\documentclass[epsfig]{aa}
\usepackage{epsfig,deluxe}
\begin{document}

\newcommand{\gsim}{\hbox{\rlap{$^>$}$_\sim$}}
  \thesaurus{06;  19.63.1}
%
\authorrunning{A. Dar \& A. De R\'ujula}
\titlerunning{A CB model of GRBs: X-ray lines}
\title{The Cannonball Model of Gamma Ray Bursts:
Lines in the
X-Ray Afterglow} 

\author{Arnon Dar$^{1,2}$ and A. De R\'ujula$^1$}
\institute{1. Theory Division, CERN, CH-1211 Geneva 23, Switzerland\\ 
           2. Physics Department and Space Research Institute, Technion,
              Haifa 32000, Israel } 
\maketitle

\begin{abstract} 

Recent observations suggest that gamma-ray bursts (GRBs) and their
afterglows are produced by jets of highly relativistic cannonballs (CBs),
emitted in supernova explosions. The fully ionized CBs cool to
a temperature below 4500 K within a day or two, at which point
electron--proton
recombination produces an intense Lyman-$\alpha$ emission.
The line energy is Doppler-shifted by the CBs' motion to  X-ray
energies in the observer's frame.  The measured line energies, corrected 
for their cosmological redshift, 
 imply Doppler factors in the range 600 to 1000, consistent with 
those estimated ---in the CB model---
from the characteristics of the $\gamma$-ray bursts. All other
observed properties of the lines are also well described by the CB model. 
Scattering and self-absorption of the recombination lines within the CB
also produce a wide-band flare-up in the GRB afterglow,
as the observations indicate.  
A very specific prediction of the CB model is that the X-ray lines
ought to be narrow and move towards lower line energies
as they are observed: their current apparently large widths would 
be the effect of time integration, and/or of the blending of lines
from CBs with different Doppler factors.

\end{abstract} 


\section{Introduction}

X-ray lines with energies of a few keV,
with very large flux and equivalent width, have been
observed in the early-afterglow phase of gamma-ray bursts
(GRBs), by the satellites BeppoSAX (GRB 970508, Piro et al. 1998; 
GRB 000214, Antonelli et al. 2000), 
ASCA (GRB 970828, Yoshida et al. 1999) and Chandra
(GRB 991216, Piro et al. 2000a). In spite of their large intensities,
these lines are near the detection limits, making their empirical
study difficult, and their interpretation debatable.

X-ray lines of similar energies, emitted
by various astrophysical systems, are
usually interpreted as Fe emission lines. Iron
X-ray lines from a GRB environment have been
discussed in the hypernova GRB scenario by Ghisellini et al. (1999)
and by Boettcher et al. (1999). But it has been argued (e.g. Lazzati et
al. 1999) that the corresponding 
 models cannot produce strong Fe-line emission and
that in the usual surrounding of GRB progenitors 
---star-formation regions--- a putative Fe-line
emission strong enough to be detectable 
should last for years, rather than for a day or two, as the observations
indicate. These difficulties were claimed not to be
present (Lazzati et al. 1999; Vietri et al. 1999; Piro et al. 2000a) if a
special geometry of the GRB circumstellar material is assumed: a torus 
perpendicular to the line of sight, 
of radius $\sim 10^{16}$ cm and density in excess of 
$\rm 10^{10}\, cm^{-3}$, made of a few solar masses of 
iron-rich material.
Such a torus might be produced in the supranova model of
GRBs (Vietri
and Stella 1998), though the model still faces many difficulties
(Vietri et al. 2001). 

The X-ray lines observed in the afterglows of GRB 970508 and 991216
can, within large errors, be interpreted as Fe lines. 
In the case of GRB 970828, the observed X-ray line energy
(Yoshida el al. 1999) corresponds to $\rm z\sim 0.38$ for a Ly$_\alpha$
line from fully-ionized Fe (or a smaller $\rm z$ for partial ionization), while the
measured redshift is $\rm z=0.957$. If the line is interpreted as the
9.23 keV recombination edge of fully-ionized Fe, the redshift is 0.84
(or smaller for incomplete ionization). But, surprisingly, no
K$_\alpha$ line was observed between 3.25 and 3.5 keV,
the range, at $\rm z=0.84$, expected for singly-ionized to fully-ionized Fe.
No independent measurement of redshift is available for GRB 000214.
All in all, the Fe-line interpretation is not established.

Here we offer an alternative interpretation of  the X-ray emission lines
in the GRB afterglows, an inevitable sequitur in 
the cannonball (CB)
model of GRBs (Dar and De R\'ujula 2000a,b). In the CB model, GRBs are
produced by highly relativistic ``cannonballs'', jetted
by nascent or dying compact stellar objects. The CBs heat up as they
pierce through nearby circumstellar shells, and their radiation is 
Lorentz-boosted to $\gamma$-ray energies by the CBs' fast motion.
Later (after a day or two in the observer's frame) 
the CBs cool sufficiently, by expansion and radiation, for their
constituent electrons and protons to recombine, producing 
a strong Ly$_\alpha$ line emission, Doppler-shifted to X-ray  
energies in the observer's frame\footnote{
Analogous Doppler-shifted Ly$_\alpha$ and
K$_\alpha$ recombination lines have been detected from the relativistic
jets of SS 433 (e.g. Margon 1984; Kotani et al. 1996).}.
We show that this interpretation is
consistent with the GRB observations and that the CB model can
correctly explain all the observed features of the lines.  
Our main prediction is that the lines should be narrow and
move in time from higher to lower frequency, their currently observed
large widths being the effect of integrating over this migration in time,
or of the blending of the emission from unresolved CBs with different 
Doppler factors. We also argue that self-absorption 
and scattering of the recombination lines in the CB produce a multiband 
flare-up of the afterglow, and we discuss the observational evidence     
for such a flare in the early afterglow of GRBs.

\section{The Cannonball Model of GRBs} 

\subsection{The engine}

In the cannonball model of GRBs (Dar and De R\'ujula 2000a,b),
we assume that large and sudden mass-accretion episodes 
onto  compact stellar objects lead to the abrupt ejection of 
highly relativistic oppositely-directed pairs
of CBs, as observed in microquasars (see, for instance, 
Mirabel and Rodriguez 1994,
1999a,b, Belloni et al. 1997; Rodriguez and Mirabel 
1999 and references therein). Such mass-accretion
episodes can take place in single-bang supernovae (Shaviv and Dar 1995; Dar 
1998; Dar and Plaga 1999; Cen 1999), 
in double-bang supernovae (De R\'ujula 1987; Woosley 1993,
Woosley and MacFadyen 1999; MacFadyen and Woosley 1999; MacFadyen et al.
1999; Dar and De R\'ujula 2000a,b),
in binaries including a compact object, in mergers of neutron stars
with neutron stars or black holes (Paczynski 1986, Goodman et al. 1987), in
transitions of neutron stars to hyperon- or quark-stars (Dar 1999; Dar and
De R\'ujula, 2000c), etc. 

The correspondance in time and location between
GRB 980425 and SN 1998bw establishes their
association quite clearly: the chance probability 
for the coincidence is less than $10^{-4}$
(e.g. Galama et al. 1998), or much smaller if the revised BeppoSAX position
(e.g. Pian et al.1999) is used in the estimate. Assuming that SN 1998bw
is close to a standard candle ---for SNe associated with GRBs---
it is straightforward to derive what its light curve as a function of
time and frequency would be, if viewed at a different
redshift. Of the sixteen GRBs with known
redshift, four or five have a modulation of their afterglow, about one
month after the GRB signal, that is very well described by the
addition of such a SN 1998bw-like signal. In all other cases there
is a good reason why such an effect would be unobservable,
e.g. the afterglow or the background galaxy are too luminous,
the data on the late afterglow are unavailable, or the measured frequency
range in SN 1998bw is insufficient to extrapolate to the required redshift.
The conclusion that more than 20\% of GRBs
are associated with SNe, and perhaps {\it all} of them, 
appears to be inescapable  (Dar and De R\'ujula 2000a). 
This is correct at least for 
the long-duration GRBs for which it has been possible
to establish the existence of an afterglow. It is conceivable that
the short-duration GRBs are not associated with (core-collapse) SNe,
but with the other stellar transitions referred to in the previous paragraph.
In what follows we concentrate on the CB model for GRBs associated
with SNe, the ones for which we have enough information to
predict the properties of X-ray lines and flares in their
afterglow.

In the CB model the jetted CBs would generate
GRBs by hitting ejecta, a stellar wind, or an envelope.
Each CB corresponds to a single pulse in a GRB light curve.
The average observed number of significant GRB pulses is
$\cal{O}\,$(5 to 10), so that a ``jet'' of CBs would typically carry that many
times the energy of a single CB. After a few CBs are emitted
by the rather random and currently unpredictable accretion-emission
process, the supply of accreting material is exhausted and the
$\gamma$-ray activity ceases (Dar and De R\'ujula, 2000a,b).

The kinetic energy of the CBs has been estimated  
from the assumption that momentum imbalance between the
opposite-direction jets in SN events is responsible for the observed large 
peculiar velocities of the resulting neutron stars: 
${\rm v_{NS}\approx 450\pm 90~ km~s^{-1}}$ (Lyne and Lorimer 1994).
Such natal kick velocities imply   $\rm E_{jet}\sim 10^{53}$ erg, or
$\rm E_{CB}\sim 10^{52}$ erg for a typical jet consisting
of $\sim 10$ cannonballs (Dar and Plaga 1999, Dar and De R\'ujula 2000a,b).
For this choice, the CB's mass is relatively small:
${\rm \sim 1.86\, M_\otimes}$ for the Lorentz factor
$\gamma={\cal{O}}(10^3)$
which we independently estimated from the GRB rate
and the GRB-supernova association,
from the average fluence per pulse in GRBs observed by BATSE,
and from the typical duration of these pulses.
We presume 
the composition of CBs to be ``baryonic'', as it is in the jets of
SS 433, from which Ly$_\alpha$ and Fe K$_\alpha$ lines have been detected
(Margon 1984, Kotani et al. 1996), although the violence of the relativistic
jetting-process may break many nuclei into their  constituents.
The baryonic number of the CB is
\begin{equation}
\rm N_b\simeq {E_{CB}\over m_p\,c^2\,\gamma}\simeq 6.7\times 10^{51}\, 
\left[{E_{CB}\over 10^{52}\, erg}\right]\, \left[{10^3\over\gamma}\right].
\label{NB}
\end{equation}
The collision of the CBs with a SN shell is so violent ---at $\sim 1$ TeV
per nucleon--- that there is no doubt that, as it exists the shell,
the CBs' baryonic number resides in individual protons and neutrons.

\subsection{The GRB}

As they hit intervening material, such as a massive shell, the CBs heat up. 
Their radiation is obscured by the shell up to a distance of order one 
radiation length from the shell's outer surface. As this point is reached, the approximately thermal surface radiation 
from the CBs ---which continue to travel, expand and cool down---
becomes visible. The radiation of a single CB, boosted and 
 forward-collimated by its
ultrarelativistic motion, and time-contracted by  
relativistic aberration (relative to the frame of an observer 
situated close to the CB's line of flight),
appears as a single pulse in a GRB.  The observed
duration of a single CB pulse is its cooling time after it becomes visible.

Approximately one third of the centre-of-mass energy 
of a CB's collision with the intervening material is converted in the
hadronic cascades within the CB into electromagnetic radiation (via $\pi^0$
production and $\pi^0\to \gamma\gamma$ decay) 
and two thirds escapes the CB as neutrinos and muons. The
enclosed electromagnetic
energy heats up the CB to a (proper) temperature in the keV domain. Let
$\gamma=\sqrt{1-\beta^2}$ be the bulk Lorentz factor of a CB with 
velocity ${\rm v=\beta\,c}$.  The Doppler factor
by which the radiation from its surface is boosted is 
given by
\begin{equation} 
{\rm \delta = {1 \over \gamma\,(1-\beta\, cos\theta)}\approx
{2\gamma\over 1+\gamma^2\theta^2}}\; ,
\label{delta}
\end{equation}
where $\theta$ is the viewing angle relative to the CB's velocity vector,
and the approximation is valid for
$\gamma \gg 1$ and $\theta\ll 1$, the case of interest here.

Let the total energy isotropically radiated  by a CB's surface, in its rest
system, be $\rm E_0$.
This radiation is boosted and collimated by the
CB's motion, its time dependence is modified by the observer's time
flowing $\rm (1+z)/\delta$ times faster than in the CB's rest system.
The fluence seen by an observer at 
redshift $\rm z$ and luminosity distance $\rm {D_L}$  is 
\begin{equation} 
\rm {d\, F\over d\,\Omega} \simeq
{E_0\, (1+z)\over 4\, \pi\, D_L^2}
 \,\delta^3\, A(\nu,z)\, , 
\label{fluence}
\end{equation} 
where ${\rm A(\nu,z)}$ is the attenuation along the line of sight.  

As long as the CB is opaque to its internal radiation, it expands in its 
rest system
at a velocity comparable to the relativistic 
speed of sound, $\rm \beta_T\,c$,
with $\rm\beta_T\,\widetilde{<}\, 1/\sqrt{3}$. Because of its relativistic 
expansion, the possibly distorted shape of a CB ---as it is
emitted or as it first collides with the SN shell--- approaches a spherical
shape at a later time; we shall treat CBs as spherical
objects and approximate their properties as if they were uniform. An
expanding CB cools quasi-adiabatically, so that
its temperature decreases as ${\rm T\propto 1/R_{CB}\propto 1/t}$. When  
a CB of mass $\rm M_{CB}\sim 2\, M_\otimes$ reaches a radius 
${\rm R_{CB}^{trans}\simeq [3\,M_{CB}\,\sigma_T/(4\, \pi\,
m_p)]^{1/2}}$ $\sim 3\times 10^{13}$ cm
(with $\rm \sigma_T\simeq 0.65\times 10^{-24}$ cm$^2$ 
the Thomson cross section) it becomes
optically thin and its remaining,
cooled-down internal radiation escapes.
This end to the CB's $\gamma$-ray pulse
takes place at $\rm t\simeq (1+z)\, R_{CB}^{trans}
/(\beta_T\,c\,\delta)$ $\rm \simeq 3\,(1+z)\,
(1/[\sqrt{3}\,\beta_T])\,(M_{CB}/M_\otimes)^{1/2}\,
(10^3/\delta)$ seconds in the observer frame.  
The spectral and temporal properties of GRB pulses are well reproduced by this
cannonball model of GRBs, as demonstrated in
detail in Dar and De R\'ujula 2000b.

\subsection{The afterglow at X-ray energies}

When the radiation enclosed in a CB escapes, its internal radiation pressure
drops abruptly and its transverse expansion rate
is slowed down by collisions with the interstellar medium.
During this phase, the CB cools mainly
by emission of bremsstrahlung and synchrotron radiation at an 
approximate rate (see, for instance, Peebles 1993)
\begin{equation}
\rm L_{brem}\simeq 1.43\times 10^{-27}\, n_e^2\, T^{1/2}\, erg\, cm^{-3}\, 
s^{-1}\, ,
\label{Lbrem}
\end{equation}
where, here and in what follows, $\rm n_e$ is in c.g.s. units and $\rm T$ is in 
degrees Kelvin.
The above cooling rate corresponds, in the observer's frame, to a cooling time
of the CB to temperature T 
\begin{equation}
\rm t_{brem}\simeq {2.9\times 10^{11}\,(1+z)\,
T^{1/2}\over n_e\,\delta}\,s\, .
\label{tbrem}
\end{equation}

Let $\rm n$  be the baryon 
number density in the CB. The  fractional ionization, $\rm x\equiv n_e/n$,
approximately satisfies the equilibrium Saha equation
\begin{equation}
{\rm {x^2 \over 1-x} = {(2\, \pi\, m_e\, c^2\, k\,T)^{3/2}\over
                     n\, h^3\, c^3}\, e^{-B/k\,T}}\, ,
\label{Saha}
\end{equation}
where $\rm B=13.6$ eV is the hydrogen binding energy.  

Given that the X-ray lines are observed to
appear in the afterglow after  $\sim 1$ day, we can use 
Eqs.~\ref{tbrem} and \ref{Saha} to estimate
the electron density at recombination time ($\rm x\sim 1/2$) to be
\begin{equation}
\rm 10^5\, cm^{-3} <n_e<10^6\, cm^{-3}\, .
\label{ne}
\end{equation}
 For electron densities in this range, the exponential term in the Saha
equation circumscribes the CB's recombination phase to a temperature
around 4500 K. 
The radius of the CB at peak recombination ($\rm x=1/2$) can be estimated
from the baryon number of Eq.~\ref{NB} and the electron density
of Eq.~\ref{ne} to be
\begin{equation}
\rm R_{CB}^{rec}=\left[{3\,N_b\over 8\,\pi\,n_e}\right]^{1\over 3}\sim
(1\, to\, 2)\times10^{15}\; cm,
\label{Rrec}
\end{equation}
a result that the cubic root makes relatively insensitive to the
input parameters. After
recombination, bremsstrahlung emission from the CB drops drastically and
bremsstrahlung and synchrotron emission from the swept up 
interstellar medium (ISM) take over, as we proceed to discuss.

Far from their source, the CBs are slowed down by the ISM they sweep, 
which has been previously ionized by the
forward-beamed CB radiation (travelling essentially at $\rm v=c$, the CB
is ``catching up'' with this radiation, so that the ISM has no time to
recombine). As in the jets and lobes of quasars, a fraction of the
swept-up ionized particles are ``Fermi-accelerated'' to cosmic-ray
energies and confined to the CB by its turbulent magnetic field,
maintained by the same confined cosmic rays (Dar and Plaga
1999). The synchrotron emission from the accelerated electrons, boosted by
the relativistic bulk motion of the CB, produces afterglows in all bands
between radio and X-rays, collimated within an angle $\rm\sim 1/\gamma(t)$
that widens as the ISM decelerates the CB and its Lorentz factor
$\rm \gamma(t)$ diminishes. ``Late'' X-ray afterglows have been
observed up to a week after the burst, after which their
flux becomes too weak to be detected (e.g., Costa 1999).

A CB of roughly constant cross section, moving in a
previously ionized ISM of roughly constant density, 
would lose momentum at a roughly constant rate, 
independently of whether the ISM
constituents are rescattered isotropically
in the CB's rest frame, or their mass is added to that of the CB.
The pace of CB slowdown is 
${\rm d\gamma/dx=-\gamma^2/x_{0}}$, with ${\rm x_{0}=M_{CB}/
(\pi\, R_{CB}^2\, n_{_{ISM}}\, m_p})$ and $\rm n_{_{ISM}}$ the number density along
the CB trajectory. 
For $\gamma^2\gg 1$,
the relation between the length of
travel $\rm dx$ and the (red-shifted, relativistically aberrant) time of 
an observer at a small angle $\theta$ is 
${\rm dx=[2\, c\, \gamma^2/(1+\theta^2\,\gamma^2)]\,[dt/(1+z)]}$.
Inserting this into $\rm d\gamma/dx$ and integrating, we obtain:
\begin{equation}
{\rm {1+3\,\theta^2\gamma^2\over 3\,\gamma^3}=
{1+3\,\theta^2\gamma_0^2\over 3\,\gamma_0^3}+
{2\,c\, t\over (1+z)\, x_{0}}}\; ,
\label{gammat}
\end{equation}
where $\gamma_0$ is the Lorentz factor of the CB as it exits
the SN shell.
The real root $\rm \gamma=\gamma(t)$ of the cubic Eq.~\ref{gammat}
describes the CB slowdown with observer's time.

In the rest frame of the CB, the particles of the ionized ISM impinge on it with
energy $\rm \gamma\, m\,c^2$. If the CB is host to a turbulent magnetic field,
the incoming particles are scattered and Fermi-accelerated 
to a power-law energy distribution,  their bremsstrahlung and 
synchrotron radiation powering the CB's late afterglow.
It is natural to assume that in a constant-density ISM 
a quasi-equilibrium is reached between the number of ISM particles entering 
and escaping the CB, so that its radiation is approximately constant 
in the CB's rest frame. This ought to
be a good approximation only for frequencies at which
the CB is transparent to its own radiation: absorption
should be relevant at the lower (radio) end of the observable spectrum.
The spectral shape of the emitted synchrotron radiation 
is ${\rm F_0\propto \nu_0^{-\alpha}}$, with
${\rm \alpha=(p-1)/2}$ and p the spectral index of the electrons. For
equilibrium between Fermi acceleration and synchrotron and Compton
cooling, $\rm p \approx 3.2$ and $\alpha\approx 1.1$, while for small
cooling rates, $\rm p \approx 2.2$ and $\alpha \approx 0.6$ (Dar and De
R\'ujula 2000b), or $\rm p\simeq1.2$ and $\alpha\simeq 0.1$ if Coulomb 
losses dominate. 
At very low radio frequencies, self-absorption becomes important and
$\alpha \approx -1/3~(2.1)$ for optically thin (thick) CBs.
For a detailed modelling of a very similar phenomenon: 
synchrotron radiation from
quasar lobes, see, for instance, Meisenheimer et al. (1989). 

The time-dependence of $\rm F_0$ should be of the form $\rm t^{-\beta}$.
For steady emission, i.e. for a small energy-deposition rate by the ISM
particles and a correspondingly small emission rate,
$\beta\simeq 0$. For an emission rate which is in equilibrium with
the energy-deposition rate in the CB rest
frame  (which is proportional to $\gamma^2$), 
$\beta\simeq 2$. All in all, the CB-model's prediction for the
``quasi-steady'' state of the CB's evolution
---from recombination to a final stopping  ``Sedov--Taylor'' phase---
is that the energy density flux in the observer's frame is approximately
of the form
\begin{equation}
{\rm F[\nu,t] \propto  
\left[{2\gamma(t)\over 1+[\gamma(t)\,\theta]^2}\right]^   
{3+\alpha+\beta}\,\nu^{-\alpha} \, A(\nu,z)}\, ,
\label{fglow}
\end{equation}  
where ${\rm A(\nu,z)}$, as in Eq.~\ref{fluence},
is the attenuation along the line of sight,
and $\rm \gamma(t)$ is the solution to Eq.~\ref{gammat}. In Eq.~\ref{fglow}
we have used Eq.~\ref{delta} for $\delta$, to expose how the
flux is forward-collimated in a ``beaming cone'' of opening angle
$\rm 1/\gamma(t)$. 

The temporal behaviour of the GRB afterglow 
of Eq.~\ref{fglow} depends on the relation
between the viewing angle $\theta$ and the initial Lorentz factor
$\gamma_0$. If an observer is initially outside the beaming cone
($\theta\,\gamma_0\! >\! 1$), since $\rm \gamma(t)$ is
a decreasing function of time, the afterglow
would increase up to a peak time $\rm t=t_p$, for
which $\rm\theta\,\gamma (t_p)=1$, and the beaming cone has widened
to the observer's angle. Thereafter the afterglow decreases.
This is the behaviour observed, for instance, in GRB 970508
(Dar and De R\'ujula, 2000a). Beyond the peak, or for 
observers within the beaming cone from the onset of the burst,
the afterglow's flux is a monotonously decreasing function of time.
For $\gamma^2\theta^2\gg 1$ and $\rm \gamma\propto t^{-1/3}$
(the asymptotic behaviour of the solution to Eq.~\ref{gammat}), 
the afterglow declines like
\begin{equation}
\rm F[\nu,t] \propto \left [{(t_p/t)^{1/3}
           \over 1+(t_p/t)^{2/3}}\right]^{3+\alpha+\beta}\, A(\nu,z)\, ,
\label{synclate}
\end{equation}
with $0\leq\beta\leq2$ (Dar and De R\'ujula
2000a,b). For radio afterglows the absorption correction is
important, not only in the ISM and the intergalactic medium,
but also within the CB itself.

The X-ray lines and multiband flare to be discussed anon are features to be
superimposed on the smooth afterglow described by Eq.~\ref{synclate}.

\section{X-ray lines in GRB afterglows}

When the temperature
in the CB drops below $\sim 5000$ K, electrons begin to recombine with protons 
into hydrogen,  causing a strong emission    
of the $\rm Ly_{\alpha}$ line (and, possibly, a recombination edge 
above the $\rm Ly_{\infty}$ line). These features are Doppler-shifted 
by the CB's motion to X-ray energies in the observer's frame:
\begin{equation} 
\rm E_\alpha\simeq {10.2\over(1+z)}\,\left[ {\delta\over 10^3}\right]\ 
keV\, ,   
\label{alpha}
\end{equation}   
\begin{equation} 
\rm E_{\infty}\simeq {13.6\over(1+z)}\,\left[{\delta\over 10^3}\right]\, 
keV\,.   
\end{equation}   
The total number of these recombination photons is approximately
equal to the baryonic number of the CB, given by Eq.~\ref{NB},
so that the photon fluence of these
lines (the line fluence) at a luminosity distance $\rm D_L$ is
\begin{equation}
\rm {dN_{lines}\over d\,\Omega}\simeq N_b\, 
         {(1+z)^2\, \delta^2\over 4\,\pi\,D_L^2}\, .
\label{linefluence}
\end{equation}
In the CB model the X-ray lines observed in GRB afterglows
are the above Ly$_\alpha$ and/or Ly$_\infty$ lines, 
not Fe lines or Fe recombination edges from a stationary source.

The redshift, luminosity distance, $\gamma$-ray 
fluence and total $\gamma$-ray energy of the GRBs
in which lines have been observed are listed in Table I
(the quoted ``spherical energy''
is that deduced under the customary assumption
---incorrect in the CB model--- that the emission is isotropic).
In Table II we report the measured attributes of the X-ray lines and,
in Table III, the properties ensuing ---in the CB model--- from their
interpretation as the boosted Ly$_\alpha$ lines of hydrogen
recombination. The inferred Doppler
factors are consistent with those needed   to explain the intensity, 
energy spectrum and line shapes of the GRB pulses $(\delta\sim 10^3)$. 

A GRB afterglow can be produced by a single CB, or by
the unresolved afterglows
of various CBs with different Doppler factors that could also 
be formed by the merger of CBs with
somewhat different velocities.   
In fact, the 4.4 keV line in the afterglow of GRB 991216 
(Piro et al. 2000a) may be either
a hydrogen recombination edge from the same CB that produces the 3.49 keV
line, or a $\rm Ly_\alpha$ line from a different CB.  Similarly,
the peak around 2 keV in the spectrum of GRB 990214 (see, for example, Fig. 
2. in Antonelli et al. 2000) may be a Doppler-shifted  $\rm K_\alpha$ 
line from another CB with a smaller Doppler factor than that of the CB
that produced the 4.7 keV line. Clear assignments will require more
precise data.

\subsection{Consistency Cheks and Predictions}

There are various independent consistency checks and testable 
predictions of the CB-model's
interpretation of the X-ray  lines as hydrogen-recombination 
features that are Doppler-shifted to X-ray energies by the  
ultrarelativistic motion of the CBs:
\vskip .2cm
\noindent
{\bf a. Time and duration of the line emission}:
The radiative recombination rate in a hydrogenic plasma to the ground state is
$\rm r_{gs}\approx 2.07\times 10^{-11}\,T^{-1/2}\, n_e\, s^{-1}$;
to any atomic level it
is $\rm r_{rec}\approx 2.52\times 10^{-10}\,T^{-0.7}\, n_e\, s^{-1}$ 
(Osterbrock 1989).
In the observer's frame the recombination time is
\begin{equation}
\rm \Delta t_{rec}\approx{4\times 10^9\,(1+z)\,T^{0.7}
\over n_e\,\delta}\; s\,.
\label{tline}
\end{equation}
At $\rm T\sim 4500$ K
the ratio of $\rm \Delta t_{rec}$ to the appearance time of the
X-ray lines, Eq.~\ref{tbrem}, is
 $\rm\Delta t_{rec}/t_{brem}\sim 0.07$, insensitive to $\rm T$
and independent of $\delta$, $\rm n_e$ and $\rm z$.

The emission time of $\rm {Ly_\alpha}$ photons, 
$\rm \Delta t_{Ly_\alpha}$, is longer than 
$\rm \Delta t_{rec}$, since the CB, at recombination time,
is not transparent to them.
Their photo-absorption cross section in neutral hydrogen is
\begin{equation}
\rm\sigma_{Ly_\alpha}\approx {\pi\, e^2\, \lambda \over m_e\, c^2} \;
f_{1,2}\simeq 2.1\times 10^{-18}\; cm^2\, ,
\label{sigmaabs}
\end{equation}
where $\lambda=1216$ \AA,
and the oscillator strength for the
$\rm n =1\rightarrow 2$ transition is ${\rm f_{1,2}=0.194}$.
At peak recombination ($\rm x\simeq 1/2$)  the
optical thickness of the CB to the Ly$_\alpha$ line is
\begin{equation}
\rm \tau=\sigma_{Ly_\alpha}\,n_e\, R_{CB}^{rec}\sim {\cal{O}}\,(10^3)\; ,
\end{equation}
with $\rm R_{CB}^{rec}$ as in Eq.~\ref{Rrec}.
Since $\tau\gg 1$, the Lyman photons are reabsorbed 
and re-emitted many times 
by the recombined hydrogen atoms in the CB  
before they escape from its surface. The duration of their 
emission is not  the recombination time, but
their diffusive residence time. In the observer's frame this time is roughly
\begin{eqnarray} 
&&\rm \Delta  t_{Ly_\alpha}\simeq (1+z)\,{3\,R_{CB}\,\tau\over c\,\delta}
\nonumber \\
&&\rm\simeq 
     10^5\,(1+z)\,\left[{R_{CB}\over 10^{15}\, cm}\right]\, 
     \left[{\tau \over 10^3}\right]\,\left[{10^3 \over \delta}\right]\, s.
\label{time}
\end{eqnarray}
This prediction is consistent with the observation that the line 
emission extended over a day or two in the early afterglows of GRB 970508 
(Piro et
al. 1999) and GRB 000214 (Antonelli et al. 2000) observed by BeppoSAX, 
and of GRB 991216 (Piro et al.  2000a) observed by Chandra. It is, however, 
inconsistent with the brevity
($\rm\Delta_t \sim 5000\,s $) of the line emission observed in GRB 970828 
(Yoshida et al. 1999), for which we have no a-priori explanation.

The above estimates are valid only for an afterglow
made by a single CB (either originally
ejected, or the result of mergers). In practice, the
afterglows of different CBs in a GRB add up, and cannot be
resolved either spatially or temporally.
Because of the possibly differing Lorentz and
Doppler factors of a succession of CBs, the observed
X-ray line emission may extend over a time longer than
the expectation of Eq.~\ref{time}. Comparison with
the data is also hampered by the fact that, in some cases,
 only a lower limit
to the duration of the line emission is available.
In the case of GRB 991216, for instance, the
gamma-ray light curve, as measured by BATSE (see, for instance, Mallozzi 2000)
has eight peaks or more,
and the line emission was detected by Chandra only during
a limited period of 30 ks, beginning 37 hours after the burst (Piro et al. 2000a).

\vskip .2cm
\noindent
{\bf b. Photon fluences:}
In the three cases for which the redshift to the GRB is known, the observed
line energies can be converted into values of the Doppler factor
$\delta$ using Eq.~\ref{alpha}, as we did in Table III. 
The resulting values of $\delta$ agree with those required in the CB
model to explain the properties of GRBs and their smooth afterglows.
The line fluence of Eq.~\ref{linefluence} can be used to estimate
the total baryon number in a CB or the
 ensemble of CBs constituting the jet of a GRB.
Since the duration $\rm \Delta t_{obs}$ 
of the available observations of X-ray lines 
around a ``central'' time $\rm t_{obs}$
may have been shorter than the actual span
$\rm \Delta t_{Ly\alpha}$ of the line emission, we can only
bound $\rm N_b$ to an interval corresponding to two extreme assumptions:
(a) $\rm \Delta t_{Ly\alpha}\sim \Delta t_{obs}$ 
and  (b) $\rm \Delta t_{Ly\alpha}\sim  t_{obs}$.
The inferred baryon numbers, listed in Table III, are within 
the expected range of Eq.~\ref{NB}.

\vskip .2cm
\noindent 
{\bf c. Line widths:} The thermal broadening of  the
recombination lines and edge is a rather small effect. But
the Doppler factors of the CBs decrease with time during the line emission,
owing to the decrease of their Lorentz factors as a result of their 
deceleration by the ISM. We have seen that sufficiently
late into an afterglow
$\rm \gamma \propto t^{-1/3}$. Thus, 
the line energy is expected to
shift by an amount 
\begin{equation}
\rm \Delta E_L \simeq E_L\,
{\Delta t_{obs}\over 3\,t_{obs}}
\label{lineshift}
\end{equation}
 during the observational interval
$\rm \Delta t_{obs}$ around $\rm t_{obs}$. When
integrated over $\rm\Delta t_{obs}$ to exploit the full statistics, the 
predicted line shift will appear as a line broadening. 
The line drifts, $\rm \Delta E_L$, estimated via Eq.~\ref{lineshift},
are listed in Table III.
They are completely
consistent with the observed widths, listed in Table II.

A line-broadening may also be caused by the blending of 
lines from CBs with similar Doppler factors, which is unresolvable 
in the afterglow phase even if the CBs did not merge 
into a single CB before e--p recombination.
Independently of the origin of the line broadening,
the predicted decrease in the line energy with time 
is a clear fingerprint of the origin of the lines: it may
be a decisive signature distinguishing the CB-model's
recombination lines from the usually assumed Fe lines.

\vskip .2cm
\noindent 
{\bf d. The CB's radius:} In Table III we list
the radii of the CBs for the different GRBs,
 approximately one observer day after the $\gamma$-ray activity.
The figures are
deduced from Eq.~\ref{Rrec} using the baryon-number values
``measured'' from the line fluence, via Eq.~\ref{linefluence}.
After one observer day, the CB is already
$\sim 1$ kpc away from the GRB emitter: far from the galactic
disk, barring a chance emission in the galactic plane.
If a CB continues to expand during the first few observer's weeks
with the same mean speed as in the first day or two,
as expected for a CB moving in a low-density galactic halo,
its radius after a month increases by a factor $\sim 30$  and 
should reach a value $\rm R_{CB}[1m] \simeq$ (3 to 6) $\times 10^{16}$ cm.
Indeed, from VLA observations of scintillations (Goodman et al. 1987) in the
radio afterglow of GRB 970508 and their disappearance after a month,
Taylor et al. (1997) inferred 
 that the linear size, $\rm 2\, R_{CB}[1m]$, of its source one month
after the burst was $\rm \approx 10^{17}$ cm. This corresponds to $\rm
R_{CB}[1m]\sim 5\times 10^{16}$ cm, in agreement with the expectation.

\section{Multiband flare up of the afterglow}

The decay of excited hydrogen atoms ---produced by recombination
or by reabsorption of emission lines--- results in a forest of
hydrogen emission lines above the Balmer limit.
These
lines are also shifted by the Doppler factor and the cosmological redshift.  
This quasi-continuous emission, which accompanies the e--p 
recombination, appears as a multiband flare in the afterglow, initially
in all wavelengths longer than $\rm 3.65\,(1+z)\, (10^3/\delta)$ \AA.
This minimal wavelength increases with time, as $\delta$ decreases. 

The total energy in the flare  is  about 
50\% of the total recombination energy, 
$\rm E_r\sim N_b\,B/2$ $\rm\sim 7\times 10^{40\pm 1}$ erg 
in the CB's rest frame, or 
$\rm \sim 7\times 10^{49\pm 1}\, (\delta/10^3)^3/(1+z)$
erg equivalent ``isotropic'' energy in the observer frame. 
For the only GRB for which a measured total energy in the
X-ray afterglow is available (GRB 970508,  Piro et al. 1998) 
the result\footnote{The reported afterglow energy
in the 2--10 keV domain is 20\% of the 
$6.6\times 10^{51}$ erg GRB energy.
The flare energy was $\sim 5\%$ of the integrated X-ray afterglow
energy.} is 
$\sim 6.6\times 10^{49}$ erg,
in agreement with the expectation.
The total fluence
of the forward-collimated and Lorentz-boosted
flare of the CB model is given approximately by
\begin{equation}
{\rm {d\, F\over d\,\Omega} \simeq
{N_b\,B\, (1+z)\over 8\, \pi\, D_L^2} \,\delta^3\, 
A(\nu,z)}\, . 
\label{flare}
\end{equation} 
This recombination flare may explain the ones observed in the 
afterglow of, for example, GRB 970228 (Masetti et al.
1997); GRB 970508 (Piro et al. 1998); GRB 990123 (Kulkarni et al. 1999);
GRB 000301c (Sagar et al.  2000) and perhaps  GRB 000926
(Veillet 2000, Piro et al. 2000b,c).
The prediction of the spectral and
temporal behaviour of the flares
requires a detailed modelling of the recombination phase in
the expanding and cooling CBs, which we have not performed.

\section{Conclusions}

We have argued that
GRBs and their afterglows may be produced
by jets of extremely relativistic cannonballs from the birth or death 
of compact stellar objects.  The cannonballs pierce through a massive 
shell or other interstellar material, are reheated by their collision 
with it, and emit highly
forward-collimated radiation, which is Doppler-shifted to $\gamma$-ray
energy. Each cannonball corresponds to an individual pulse in a GRB. The CBs
decelerate by sweeping up the ionized interstellar matter in front of
them, part of which is accelerated to cosmic-ray energies and emits
synchrotron radiation: the afterglow. When the cannonballs cool below 
4500 K, electron--proton recombination to hydrogen produces Ly-$\alpha$
lines that are  Doppler-shifted to X-ray energies by the CBs' ultrarelativistic
motion. 

The properties of the  X-ray lines  
discovered so far in the afterglow of four GRBs are completely consistent  
with this picture. 
The predicted decrease in the line energy with time 
is a clear fingerprint of the origin of the lines: it may
be a decisive signature distinguishing the CB-model's
recombination lines from the usually assumed Fe lines. 
The Fe-line interpretation has the advantage of implying
a redshift value ---that can be measured independently---
and may or may not strengthen the corresponding hypothesis.
It is not inconceivable that the CB model may accommodate
Fe lines as well. Indeed, in this model, the GRB emission 
is not a ``delayed'' event, as in the 
``supranova'' model of Vietri and Stella (1998) and Piro et al. (2000a), 
and Vietri et al. (2001),
but occurs only from a few hours to a day after the SN explosion.
Very little is empirically known about the emission properties
of SNe at such an early time.

Despite their unexpectedly large intensities, the observed lines were near the
limit of the BeppoSAX, ASCA and Chandra sensitivities,
 making the study of
their precise properties quite difficult. In particular, the present
accuracy is not sufficient to test whether the line energies are redshifted
Fe lines (Piro et al. 1999) or Doppler-boosted hydrogen lines.  More
observations are needed with higher statistics and energy resolution, in
order to determine the precise energy, width and temporal evolution
of the X-ray lines, and
to establish their true identity and production mechanism. 

Part of the recombination energy absorbed in the CB produces a 
multiband flare up of the afterglow, whose energy is mainly in X-rays. 
These flares should be a common feature of the
early afterglow of GRBs. In the case of GRB 970508, the flare
energy in X-rays is measured (Piro et al. 1998)
and it agrees with the expectation.

The CB model of GRBs gives a good description of the properties
of the $\gamma$-ray light curves, and of their energy distributions
(Dar and De R\'ujula 2000b). It is also succesful in explaining the
properties of their afterglows (Dar and De R\'ujula 2000a). 
Plaga (2000) has argued that the empirical
relations between GRB variability, luminosity, resdhift and
``spike'' intensity (Fenimore and Ramirez-Ruiz, 2000) are 
consequences of the CB model. In this paper
we have shown that the model also offers an interesting and
predictive explanation of the X-ray lines and 
multiband flares in the afterglow of GRBs. 

\noindent
{\bf Acknowledgements}: We are grateful to Ehud Behar, Shri 
Kulkarni, Ari Laor and Rainer Plaga  for useful 
discussions and suggestions. This research was supported
in part by the Fund for Promotion of Research 
and the Hellen Asher Space Research Fund at the Technion.

{}

\newpage
\vskip 0.3 true cm
{\bf
\noindent
Table I -  GRB afterglows with X-ray lines}
\begin{table}[h]
\hspace{.8cm}
\begin{tabular}{|l|c|c|c|c|c|c|l|}
\hline
\hline
GRB   &z  &D$_{\rm L}$ &${\rm F_\gamma}$
&${\rm E_\gamma}$ & M\\
\hline
970508   &0.835     &5.70  &0.31  & 0.066      & 25.7  \\
970828   &0.957     &6.74  &7.4   & 2.06       &  ---  \\
991216   &1.020     &7.30  &25.6  & 8.07       & 24.5  \\
000214   & ---      & ---  &1.4   & ---        & ---   \\
\hline
\end{tabular}
\end{table}
\vskip -0.3 true cm
\noindent
{\bf Comments:} Redshift $\rm z$.
Luminosity distance D$_{\rm L}$ in Gpc (for
${\rm \Omega_m=0.3,
\; \Omega_\Lambda=0.7}$ and ${\rm H_0=65\, km\, s^{-1}\,Mpc^{-1}}$).
Fluences ${\rm F_\gamma}$ in the 40--700 keV band, in
$10^{-5}$ erg cm$^{-2}$ units.
Equivalent isotropic energy ${\rm E_\gamma}$ in units of
$10^{53}$ ergs.
R-magnitude $\rm M$ of the host galaxy.

\vskip 0.8 true cm
{\bf
\noindent
Table II - Observed X-Ray line properties}
\begin{table}[h]
\hspace{.2cm}
\begin{tabular}{|l|c|c|c|c|c|c|l|}
\hline
\hline
GRB  &   $\rm E_{line}$ &{\rm Width}
& $\rm t_{obs}$ & $\rm \Delta t_{obs}$
& $\rm I_{line}$  \\
\hline
970508 &   3.4 &$\leq 0.50$ & 26  & 11 &  $5.0\pm 2.0$  \\
970828 & 5.04 & $0.31^{+0.38}_{-0.31}$ & 128 &24 &  $1.9\pm 1.0$\\
991216 & 3.49 & $0.23\pm 0.07$ & 139 & 12 & $3.2\pm 0.8$\\
991216 & 4.4  & --- & 139 & 12 & $3.8\pm 2.0$ \\
\hline
\end{tabular}
\end{table}
\vskip -0.3 true cm
\noindent
{\bf Comments:} Line energy $\rm E_{line}$ and line width,
in keV.
Line intensity $\rm I_{line}$ in $\rm cm^{-2}\, s^{-1}$. 
Mean observation time $\rm t_{obs}$ and observed line emission
period $\rm \Delta t_{obs}$ in ks.

\vskip 0.8 true cm
{\bf
\noindent
Table III - CB Properties from the X-Ray lines}
\begin{table}[h]
\hspace{.7cm}
\begin{tabular}{|l|c|c|c|c|c|c|l|}
\hline
\hline
GRB  & $\rm \Delta E_L$
&$\delta$& $\rm N_b^{(a)}$&$\rm N_b^{(b)}$&
$\rm R_{CB}^{rec}$\\
\hline
970508 & 0.48& 612& $1.7$&8.5&1.0 \\
970828 &0.31 &967&$0.7$&4.0& 1.1\\
991216 &0.10& 691 & $1.3$&15.7&1.6 \\
991216 &0.13 & 800 & $1.3$&15.7&1.8\\
000214 &0.36 &$\geq$ 460 & --- & --- & --- \\
\hline
\end{tabular}
\end{table}
\vskip -0.3 true cm
\noindent
{\bf Comments:} Line drift $\rm \Delta E_L$ in keV. Doppler
factor $\delta$ for the hydrogen
Ly$_\alpha$-line interpretation. Baryon number 
$\rm N_b$ in $10^{51}$ units, for: 
(a) $\rm \Delta t_{Ly\alpha}\sim\Delta t_{obs}$
(b) $\rm \Delta t_{Ly\alpha}\sim t_{obs}$. 
CB's radius at recombination $\rm R_{CB}^{rec}$, in $\rm 10^{15}\, cm$ units.

\end{document}